\documentclass[journal = jpcbfk]{achemso}
\setkeys{acs}{usetitle = true}
\usepackage{graphicx}
\usepackage{color}
\usepackage{amsmath}
\newcommand{\rmi}{\rm i}
\newcommand{\rme}{\rm e}
\newcommand{\rmd}{\rm d}

\def\be{\begin{equation}}
\def\ee{\end{equation}}
\def\bea{\begin{eqnarray}}
\def\eea{\end{eqnarray}}

\def\bra#1{\mbox{$\langle#1|$}}
\def\ket#1{\mbox{$|#1\rangle$}}

\author{Tobias Kramer}
\affiliation{Konrad-Zuse-Zentrum f\"ur Informationstechnik Berlin, 14195 Berlin, Germany}
\alsoaffiliation{Harvard University, Department of Physics, 17 Oxford St, 02138 Cambridge, MA, USA}
\email{kramer@zib.de}
\author{Mirta Rodr\'iguez}
\affiliation{Konrad-Zuse-Zentrum f\"ur Informationstechnik Berlin, 14195 Berlin, Germany}
\author{Yaroslav Zelinskyy}
\affiliation{Konrad-Zuse-Zentrum f\"ur Informationstechnik Berlin, 14195 Berlin, Germany}
\title{Modeling of Transient Absorption Spectra in Exciton Charge-Transfer Systems}
\date{\today} 
\begin{document}

\begin{abstract}
Time-resolved spectroscopy provides the main tool for analyzing the dynamics of excitonic energy transfer in light-harvesting complexes.
To infer time-scales and effective coupling parameters from experimental data requires to develop numerical exact theoretical models.
The finite duration of the laser-molecule interactions and the reorganization process during the exciton migration affect the location and strength of spectroscopic signals.
We show that the non-perturbative hierarchical equations of motion (HEOM) method captures these processes in a model exciton system, including the charge transfer state.
\end{abstract}

\section{Introduction}

The understanding of electron and charge transfer mechanisms in photosynthetic systems is key for designing novel artificial photofunctional devices and to the synthesis of new functional electronic materials \cite{Curutchet2016}.
Charge transfer dynamics in the reaction center of light harvesting complexes has been probed using different spectroscopic techniques \cite{Visser1995, Klug1995, Greenfield1999, Prokhorenko2000a,Groot2005,Myers2010}.

Charge separation in photosystems is not fully understood \cite{Reimers2016a} and requires the development of new theoretical methods and computational tools for an unified description of the exciton and charge transfer phenomena.
Different theoretical approaches to describe the electron transfer have been proposed based on treating the coupled electronic and nuclear degrees of freedoms as a dissipative quantum system \cite{Okada1999}. 
There exist different treatments with respect to the vibrational state of the constituent pigments, either by a common heat bath \cite{Tanaka2010}, or by assigning distinct heat baths to the constituents \cite{Fuller2014}. In the latter approach, charge transfer (CT) states are treated on a similar footing as the Frenkel excitons in the system 
to describe the experimental spectroscopic results for reaction centers of light harvesting complexes \cite{Renger2002,Raszewski2005,Novoderezhkin2007a,Shibata2013,Lewis2013}.

The interpretation of experimental data in terms of inter- and intramolecular processes requires to fit the site energies and molecular couplings of model parameters to the time-resolved spectra.
A common parametrization of excitonic energy transfer is based on the Frenkel exciton description \cite{May2004}.
The resulting open quantum system dynamics is then often analysed within Redfield or generalized F\"orster theories. 
The applicability of Redfield and generalized F\"orster theories requires to divide the pigment complex in domains treated by the respective method.
For the light harvesting complex II (LHC II) such a combined treatment strongly affects the dynamics of the relaxation process, as a comparison with the non-perturbative hierarchical equations of motion (HEOM) method shows \cite{Kreisbeck2014}.

In addition, the exciton-CT models assign different reorganization energies to the different pigments, with the charge transfer states undergoing the largest reorganization shifts \cite{Olsina2014,Kreisbeck2015}.
In particular the dynamics towards the reaction center of photosystem II is debated with different models proposed \cite{Reimers2016a}.

The formally exact HEOM method \cite{Tanimura1989} is well suited for describing systems with reorganizational shifts, since it accounts for the finite time-scale of the reorganization process \cite{Ishizaki2009}.
HEOM facilitates the computation of time-dependent spectroscopic experiments, since it is implemented as a real-time propagation method, which allows one to directly include the time-dependent electromagnetic field. 
HEOM has been used before to describe the 2d-spectroscopic experiments of the Fenna-Matthew-Olson (FMO) complex \cite{Chen2011,Hein2012,Jing2015}.

Here, we extend the HEOM method to calculate transient absorption spectra (TA) for a model exciton-CT system.
To elucidate how a CT state and specific laser pulses affect pump-probe experiments, we put forward a minimal model to investigate with the numerically exact HEOM.
For simplicity we have not included any midway molecules mediating the electron transfer and thus do not discuss superexchange/sequential processes \cite{Bixon1991,Sumi2001}.
In contrast to previous TA simulations with HEOM \cite{Bai2015}, we include excited state absorption caused by the finite laser pulses.

The paper is organized as follows.
First, we briefly describe the Frenkel-exciton model and the light-molecule interaction within this model. 
Next, we show how the HEOM method is applied to obtain the system dynamics and to compute the transient absorption spectra.
We discuss the resulting transient absorption spectra along with an analysis of the kinetics of the exciton-CT model driven by the sequence of pump-probe laser pulses.
This allows us to connect coherence times and relaxation times of the model system to the decay of the transient signals.
In particular, the thermalization process for finite reorganization energies is properly treated by the HEOM method.
Depending on the laser parameters (pump width and frequency), the excited state absorption is enhanced.
For broadband pulses, coherences prevail in the integrated transient absorption spectra.

\section{Frenkel exciton model with light induced dynamics}\label{sec:Model}

The light harvesting complex is modeled within the Frenkel exciton description \cite{May2004}, including the external electromagnetic field from the laser pulse $H_{\rm field}(t)$,
\bea
\label{eq:h-full}
H(t)=H_{\rm g}+H_{\rm ex}+H_{\rm bath}+H_{\rm ex-bath}+H_{\rm field}(t).
\eea
Here, $H_{\rm g}=\varepsilon_0\ket{0}\bra{0}$ represents the ground state Hamiltonian (ground state energy $\varepsilon_0$), $H_{\rm ex}$ denotes the excitation energies and interactions of the pigments, and $ H_{\rm bath}$ models the effect of the surrounding protein environment as a vibrational bath coupled to each pigment by $H_{\rm ex-bath}$.  
The excitonic Hamiltonian $H_{\rm ex}$ for a system of $N$ constituents is parametrized as
\bea
\label{eq:h-ex}
H_{\rm 0} =\sum_{m=1}^{N}\varepsilon_m^0\ket{m}\bra{m}+\sum_{n\neq m}J_{mn}\ket{m}\bra{n},\quad
H_{\rm ex} =H_{\rm 0}+\sum_{m=1}^{N}\lambda_m\ket{m}\bra{m},
\eea
where we introduce the energy $\varepsilon_{m}=\varepsilon_{m}^{0}+\lambda_m$, which consists of the zero phonon energy  $\varepsilon_{m}^{0}$ shifted by the reorganization energy $\lambda_m$, and the coupling matrix elements $J_{mn}$.
The vibrational environment consists of $N$ uncorrelated baths $H_{{\rm bath},m}=\sum\limits_{i}\hbar\omega_{m,i}(b_{m,i}^{\dagger}b_{m,i}+\frac{1}{2})$ of harmonic oscillators of frequencies $\omega_{m,i}$, with bosonic creation and annihilation operators $b_{m,i}$. 
The oscillator displacement of each bath $(b^{\dagger}_{m,i}+b_{m,i})$ is coupled to the exciton-CT system by
\begin{equation}
H_{\rm ex-bath}=\sum_m \ket{m}\bra{m} \otimes \sum_i \hbar\omega_{m,i} d_{mi}(b^{\dagger}_{m,i}+b_{m,i}),
\end{equation}
where $d_{mi}$ denotes the coupling strength related to the spectral density $J_m(\omega)=\pi\sum_i \hbar^2\omega_{mi}^2d_{mi}^2\delta(\omega-\omega_i)$.
The spectral density is also connected to the reorganization energy
\bea
\lambda_m=\int_0^\infty \frac{J_m(\omega)}{\pi\omega}\rmd\omega.
\eea
Here, we consider a Drude-Lorentz shape spectral density
\bea
\label{eq:spectral_density_DL}
J_{m}(\omega)=2\frac{\lambda_{m}\omega\nu_m}{\omega^2+\nu_m^2}\;,
\eea
with inverse bath correlation time $\nu_{m}^{-1}$.

\subsection{Coupling to the laser field}
\label{subsec:coupl_laser}

Each pigment is assigned a transition dipole vector $\mathbf{d}_m$, which under irradiation with light of polarization ${\mathbf e_p}$ contributes to transitions from the ground state to the exciton states:
\begin{equation}
\hat\mu^+=\sum_{m=1}^{N}  \mathbf{d}_m\cdot {\mathbf e_p} |m\rangle\langle 0|\,.
\end{equation}
Similarly, the dipole operator for de-excitation is given by $\hat\mu^-=(\hat\mu^+)^T$.
We include multiple excitations of the system in the description by adding states representing two simultaneous excitations $S_2$ to the single excitation Hamiltonian $H_{\rm ex}$ ($S_1$). 
Ref.~\citenum{Novoderezhkin2005} considers all possible combinations of additional excitonic and CT excitations (including two excitations on the same constituent), while Ref.~\citenum{Myers2010}  excludes any CT states from $S_2$.
Here, we consider all CT and excitonic states as Fermions and exclude two excitations of the same state.
This leads to $\left[N (N - 1)\right]/2$ $S_2$ states given by Ref.~\citenum{Hein2012} Eqs.~(21,22).
The dipole moments of the $S_2$ states are given by Ref.~\citenum{Hein2012} Eq.~(23).
The transition dipole moment of the CT state is set to zero, which reduces the contribution of CT states to the absorption.
The dipole operator is multiplied by the electric field amplitude of the pump ($\rm pu$) and probe ($\rm pr$) pulse
\begin{equation}
E(\mathbf{r},t)=E_{\rm pu}(\mathbf{r},t)+E_{\rm pr}(\mathbf{r},t)\,.
\label{eq:laser_field}
\end{equation}
The pump pulse prepares a nonstationary state, which is monitored by the time-delayed weak probe pulse. The real valued electric field of the $p= \{ {\rm pu,pr} \}$ pulse is given by $E_p(\mathbf{r},t)=E^+_p(\mathbf{r},t)+E^-_p(\mathbf{r},t)$, where $E^-_p=(E^+_p)^*$ and
\bea
\label{eq:probe_pulse}
E^+_p(\mathbf{r},t)=\tilde{E}_p(t-t_{c,p})\rme^{\rmi(\omega^c_p t+\mathbf{k}_p\mathbf{r})}\,
\eea
$\tilde{E}_p(t)$ denotes the pulse envelope, centered at $t_{\rm c}$, $\omega^c$ the carrier frequency, and $\varphi=\mathbf{k}\cdot\mathbf{r}$ the phase of the laser pulse.
The delay time between pump and probe pulse is given by $\tau_{\rm del}=t_{\rm c,pr}-t_{\rm c,pu}$.
Within the rotating-wave approximation (RWA), the complex valued electric field is combined with the respective excitation and de-excitation parts of the dipole operator \cite{Gelin2009,Gelin2013,Chen2015}:
\begin{equation}
\label{eq:h-field-rwa}
H_{\rm field}(t)=-{\hat \mu}^-\big[E^+_{\rm pu}(t-t_{\rm c, pu}) +E^+_{\rm pr}(t-t_{\rm c,pr})\big] + {\rm h.c.} 
\end{equation}
The dynamics of the system driven by the external laser field is described by the Liouville-von Neumann equation for the total (system+bath) density matrix ${\hat\rho}_{\rm tot}(t)$,
\begin{equation}
\label{eq:LNE}
\frac{\partial}{\partial t}\hat{\rho}_{\rm tot}(t)=-\frac{i}{\hbar}[H(t),\hat{\rho}_{\rm tot}(t)].
\end{equation}
The physical observables are computed from the reduced density matrix $\hat{\rho}(t)$ by taking the partial trace of the full density matrix with respect to the bath modes 
\bea
\label{eq:reduced_density_matrix}
\hat{\rho}(t)={\rm tr}_{\rm bath}\{\hat{\rho}_{\rm tot}(t)\}.
\eea
We perform the time propagation using the HEOM method introduced by Tanimura and Kubo \cite{Tanimura1989}.
\subsection{Computation of transient absorption spectra}
\label{sec:ta}

Transient absorption (TA) or pump-probe spectroscopy provides a tool for measuring time-dependent phenomena in molecular networks and reflects the coherent and dissipative part of the dynamics \cite{Nuernberger2015}.
The time dependent optical response of the molecules is characterized by the laser induced polarization ${\mathbf P}(t)$. 
To study the effect of a finite laser pulse, we  utilize the non-perturbative approach developed by Seidner et al.\ \cite{Seidner1995} to calculate the non-linear polarization and the correspondent femtosecond pump-probe signal.
For weak fields and in the impulsive pulse limit, transient spectra can be obtained from the perturbative expansion of the polarization in terms of the field strength. Then the two Fourier transforms of the third order response function $S^{(3)}(\omega_3,\tau_{\rm del},\omega_1)$ and an additional integration yield the transient absorption spectra in the impulsive limit \cite{Mukamel1995,Hamm2011} at delay time $\tau_{\rm del}$
\begin{equation}
\Delta S_{\delta}(\omega_3,\tau_{\rm del})={\rm Re} \int_{-\infty}^\infty\rmd\omega_1\; S^{(3)}(\omega_3,\tau_{\rm del},\omega_1).
\end{equation}
Within the non-perturbative approach, the time-dependent polarization $P(t)$ due to the laser pulse is given by
\begin{equation}
P(t)={\rm tr}\{\hat{\rho}(t) \hat{\mu}^+\}, \quad \hat\rho(t=0)=|0\rangle\langle 0|
\label{eq:polarization}
\end{equation}
where $\hat\rho(t)$ denotes the time-evolved density matrix from the time-dependent Hamiltonian (\ref{eq:h-full}). The trace is taken with respect to the system and bath.
The energy absorption rate is proportional to the total energy dissipated by the probe pulse in the medium \cite{Mukamel1995}, eq.~(4.81),
\begin{equation}
S(t)=\frac{\rmd \langle W(t)\rangle}{\rmd t}=-P(t)\,\frac{\partial E_{\rm pr}(t)}{\partial t}\,.
\label{eq:pr_signal}
\end{equation}
If $\omega_{\rm pr}E_{\rm pr}\gg \partial E_{\rm pr}(t)/\partial t$ this simplifies with Parseval's theorem to
\begin{equation}\label{eq:dip_trans_sig}
S(\omega,\tau_{\rm del})=2\,\omega^c_{\rm pr}\,{\rm Im}[{\cal E}_{\rm pr}(\omega){\cal P}^{*}(\omega)],
\end{equation}
with the Fourier transformed polarization ${\cal P}(\omega)$ and electric field ${\cal E}(\omega)$.
For a heterodyne phase averaged detection scheme, four propagations of the initial density matrix with different phases of the pump field are required, while the phase of the probe field is set to zero \cite{Seidner1995}, (3.13a):
\begin{eqnarray}\label{eq:phaseav}
\bar{P}(t)&=\frac{1}{4}\big[
P(t,\varphi_{\rm pu}=0)+
P(t,\varphi_{\rm pu}=\frac{\pi}{2})\nonumber\\
&+P(t,\varphi_{\rm pu}=\pi)+
P(t,\varphi_{\rm pu}=\frac{3\pi}{2})\big]
\end{eqnarray}
The transient absorption spectra at a delay time $\tau_{\rm del}$ result from subtracting the spectra with and without the pump field
\begin{equation}
\label{eq:dif_absorption}
\Delta S(\omega,\tau_{\rm del})=
2\,\omega_{\rm pr}\,{\rm Im}[{\cal E}_{\rm pr}(\omega) 
(\bar{\cal P}^{*}(\omega)-{\cal P}_{\rm only~pr}^{*}(\omega))]
\end{equation}
The additional integration over all frequencies yields the integrated transient spectra
\begin{equation}
\label{eq:dif_absorption_integrated}
\Delta \tilde{S}(\tau_{\rm del})=\int_{-\infty}^{\infty}d\omega\,\Delta S(\omega,\tau_{\rm del})\,.
\end{equation}
All spectra are laser-phase averaged  (\ref{eq:phaseav}) and in addition rotationally averaged over 20 different orientations of the molecular complex with respect to the laser polarization, see Sect.~4 in Ref.~\citenum{Hein2012}.
This requires to perform 40 polarization runs to obtain the transient spectrum for a single delay time.
The HEOM implementation requires an increasing amount of memory with the number of exciton states $N$ and the truncation depth $K$ chosen to achieve numerical convergence $(N M)^{(K+2)}$, where $M$ denotes the number of  Matsubara frequencies.
The inherent parallelism of the HEOM with many coupled differential equations for the propagation of the auxiliary density operators makes HEOM well suited for acceleration on many-core computing devices such as graphics processing units \cite{Kreisbeck2011,Strumpfer2012a,Kreisbeck2014}.

\section{Exciton charge-transfer dynamics}
\label{sec:results}

We study a minimal model which incorporates some features found in the photosystem II core complex reaction center (PSII~RC) \cite{Nelson2014,Myers2010}.
Charge transfer kinetics in PSII~RC can be understood as a combination of successive charge transfer processes into four charge transfer states \cite{Novoderezhkin2011,Romero2014a}.
The model includes two pigments ($P_{D_1}$, $P_{D_2}$) of the PSII~RC at sites 1 and 2.
Charge transfer within this pair is the initial step within the core complex. 
Following Ref.~\citenum{Novoderezhkin2011}, the charge transfer state $P_{D2}^{+} P_{D1}^{-}$ is added to the Frenkel exciton model as an additional two-level system with vanishing transition dipole.
The site energies and couplings of the model Hamiltonian are taken from \cite{Novoderezhkin2011}
\begin{equation}
  H_{0}=
  \left[{
  \begin{array}{ccc}
   14720 & 150 & 45 \\    
   150 & 14650 & 45 \\
        45 & 45 & 14372          
   \end{array} }
   \right]\,\text{cm$^{-1}$.}
\label{eq:Hamiltonian-site-basis-H0}
\end{equation}
For the optical response function, we extend the one exciton Hamiltonian to the two-exciton and exciton-CT states to take into account excited state absorption caused by multiple laser pulses.
The dipole moments are taken along the axis of the nitrogen atoms N$_B$, N$_D$ in the 2AXT structure \cite{Loll2005}, resulting in an angle of 126 degrees between $P_{D_1}$ and $P_{D_2}$, while the dipole moment for the charge transfer state is set to zero.
The reaction center is set apart from the purely excitonic part of the system by its differing coupling strength to the vibrational modes \cite{Novoderezhkin2005,Gelzinis2013a}.
Following Ref.~\citenum{Novoderezhkin2011}, we increase the reorganization energy of the charge transfer state by $1.5$ compared to the core pigments (sites 1,2), $\lambda_1=\lambda_2=\lambda_3/1.5$. 
We perform calculations varying $\lambda_1=20$~cm$^{-1}$, $100$~cm$^{-1}$ and setting the damping constants in the Drude-Lorentz shaped spectral density (\ref{eq:spectral_density_DL}) to $\nu^{-1}_1=\nu^{-1}_2=\nu^{-1}_3=50$~fs.    
The resulting reorganization shift modifies the thermalization time scale by lowering the eigenvalues of the exciton system to their bare values.
This effect is contained in the HEOM method \cite{Ishizaki2009,Olsina2014}, but not within Redfield or F\"orster methods, where the finite time scale of the relaxation is not part of the theory.

\begin{figure}[t]
\includegraphics[width=0.6\textwidth]{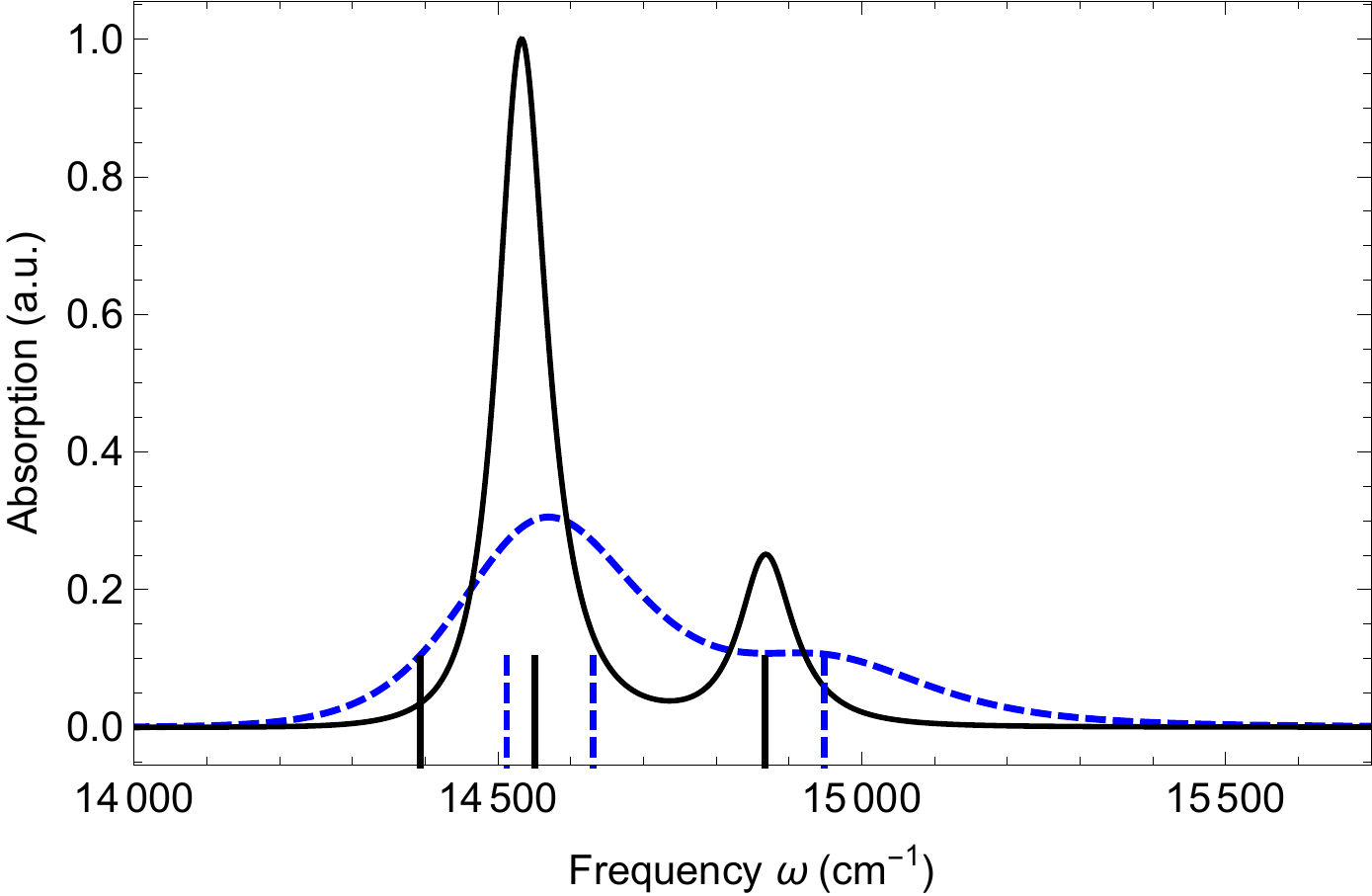}
\caption{Rotationally averaged absorption spectra of the exciton-CT model 
for weak $\lambda=\{20,20,30\}$~cm$^{-1}$ (solid line) and strong coupling $\lambda=\{100,100,150\}$~cm$^{-1}$ (dashed) to the bath at $T=300$~K.
Vertical lines show the three $H_{\rm ex}$ eigenstates, the lowest one (CT) has zero 
dipole moment.
\label{fig:abs}
}
\end{figure}

\subsection{Transient absorption spectra}

Basic spectroscopic information about the dipole moments and excitonic energy states is contained in the linear absorption spectra shown in Fig.~\ref{fig:abs} at temperature $T=300$~K, computed with HEOM using Ref.~\citenum{Hein2012}, Eq.~(12).
For the reduced three state system only two peaks are visible due to the radical pair carrying a vanishing transition dipole moment.  $P_{D_1}$ and $P_{D_2}$ are contributing to the main two peaks in the absorption spectra of the full RC system.
The relative dipole orientation leads to the biggest absorption around 14500 ~cm$^{-1}$. 
With increasing $\lambda$, the line shapes broaden and the maxima are moving. Additionally, the peak position is affected by changing the eigenvalues of $H_{\rm ex}$, while keeping $H_0$ fixed.
\begin{figure}[t]
\includegraphics[width=0.6\textwidth]{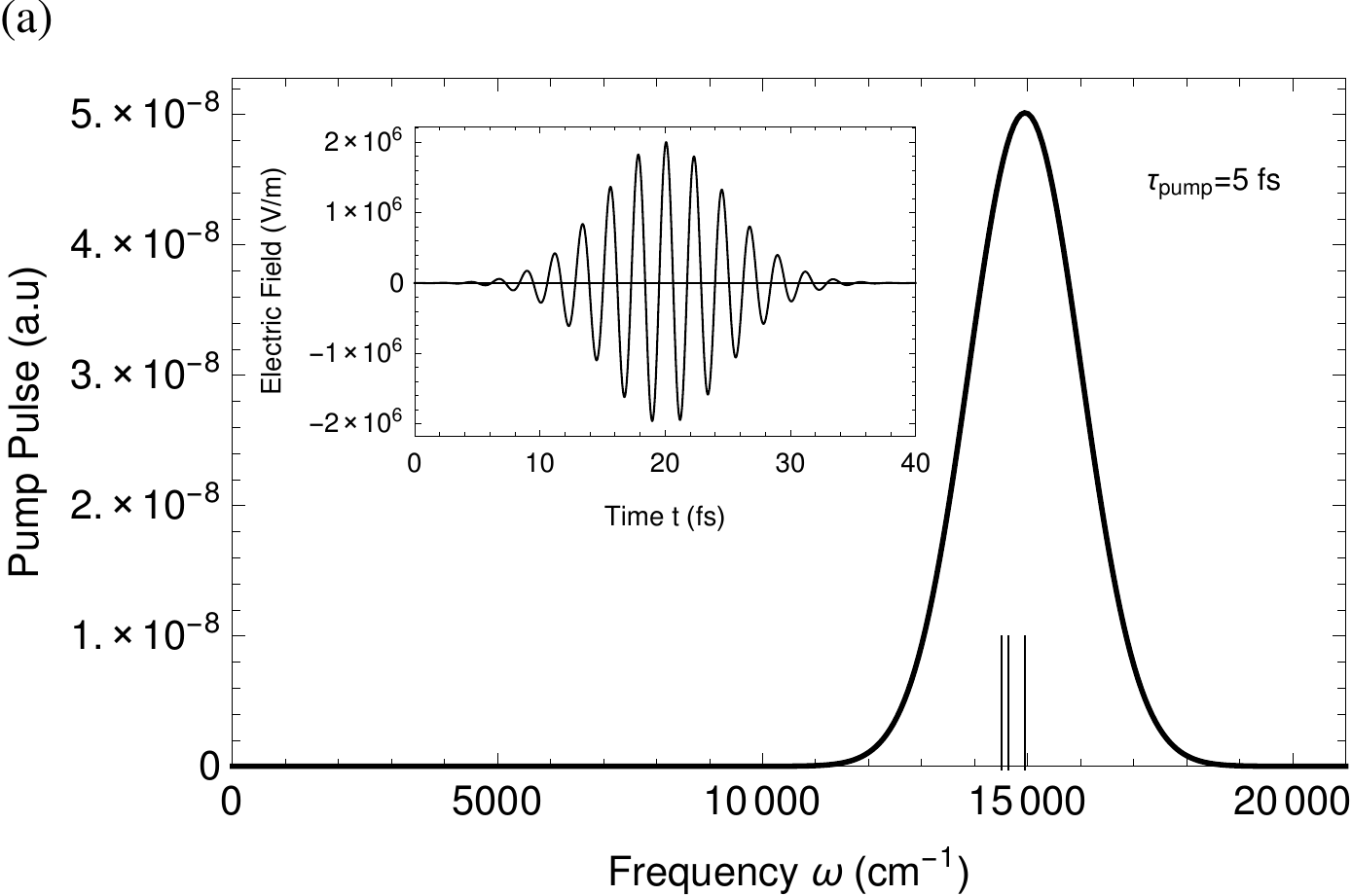}
\includegraphics[width=0.6\textwidth]{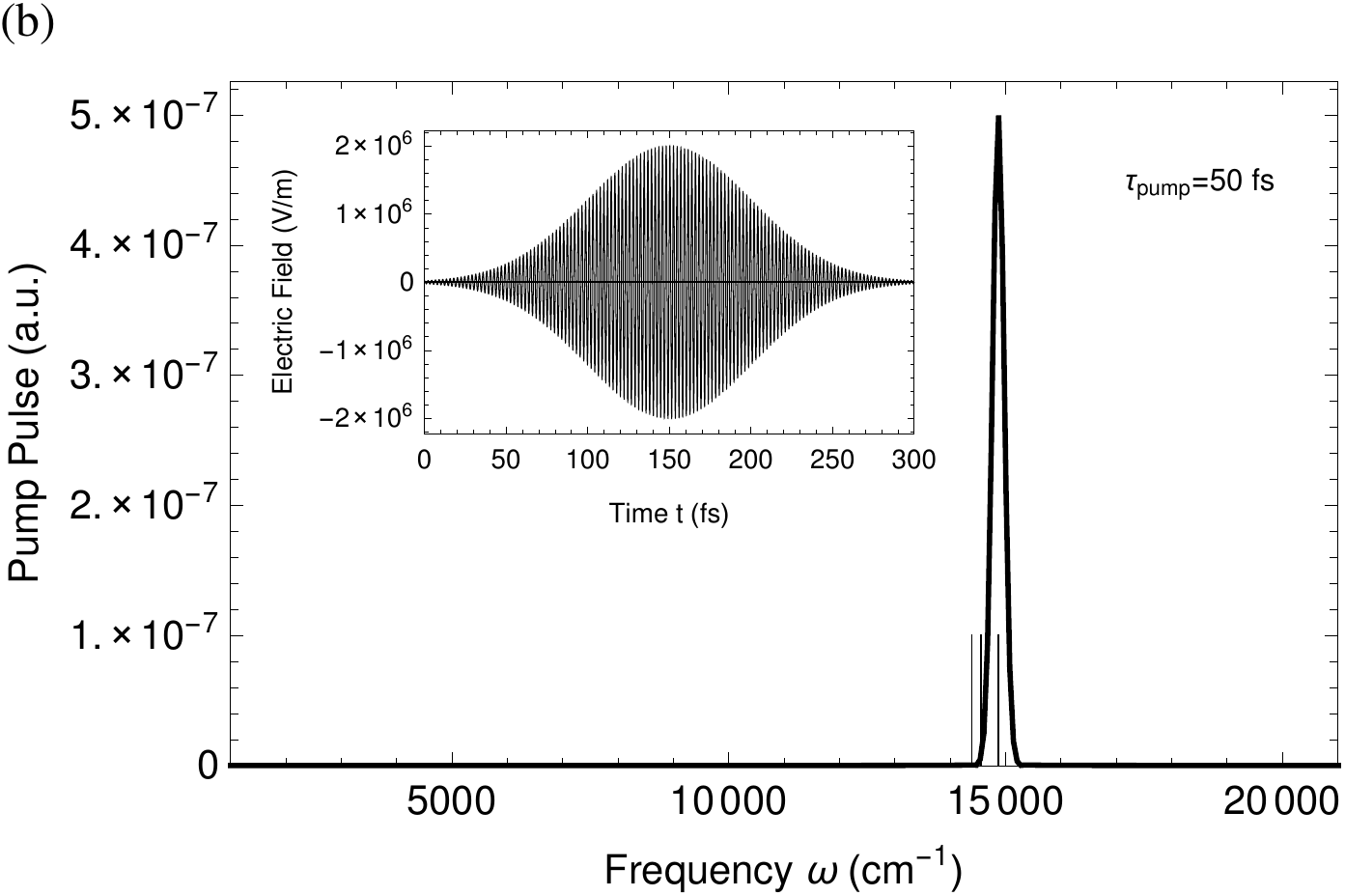}
\caption{Frequency range covered by the pump laser pulse for two different pulse widths $\tau_{\rm pu}=5$~fs (a), $50$~fs (b). The inset shows the time-dependent electric field amplitude. The vertical lines indicate the eigenvalues of $H_{\rm ex}$,  $\lambda=\{20,20,30\}$~cm$^{-1}$.}
\label{fig:EF_fourier}
\end{figure}

Next, we compute the transient absorption spectra $\Delta S (\omega,\tau_{\rm delay})$ using the time dependent coupling to the laser pulse.
We  consider a Gaussian envelop for both pump and probe pulses 
\begin{equation}
\tilde{E}_{\rm p}(t)=E^0_{\rm p}\exp(-(t-t_{\rm c,p})^2/2\tau^2_{\rm p}), \quad {\rm p}={\rm pu,pr},
\end{equation}
separated by the delay time $\tau_{\rm del}$. 
The carrier frequency of the pump pulse ($E^0_{\rm pu}=2\cdot 10^6$~V/m) is chosen to be resonant to the transition between the ground state and the highest exciton state ($\omega^c_{{\rm pu},{\lambda=20}}=14868$~cm$^{-1}$, $\omega^c_{{\rm pu},{\lambda=100}}=14948$~cm$^{-1}$).
The short probe pulse ($\tau_{\rm pr}=5$~fs, $E^0_{\rm pr}=5\cdot 10^4$~V/m) with a wide frequency dispersion is tuned to $\omega^c_{{\rm pr},{\lambda=100}}=14641$~cm$^{-1}$ and $\omega^c_{{\rm pr},{\lambda=20}}=14560$~cm$^{-1}$.
The width of the pump laser  pulse is varied from $\tau_{\rm pu}=5$~fs to  $\tau_{\rm pu}=50$~fs.
The frequency range covered by the pulses is shown in Fig.~\ref{fig:EF_fourier}.
\begin{figure}[t]
\includegraphics[width=0.6\textwidth]{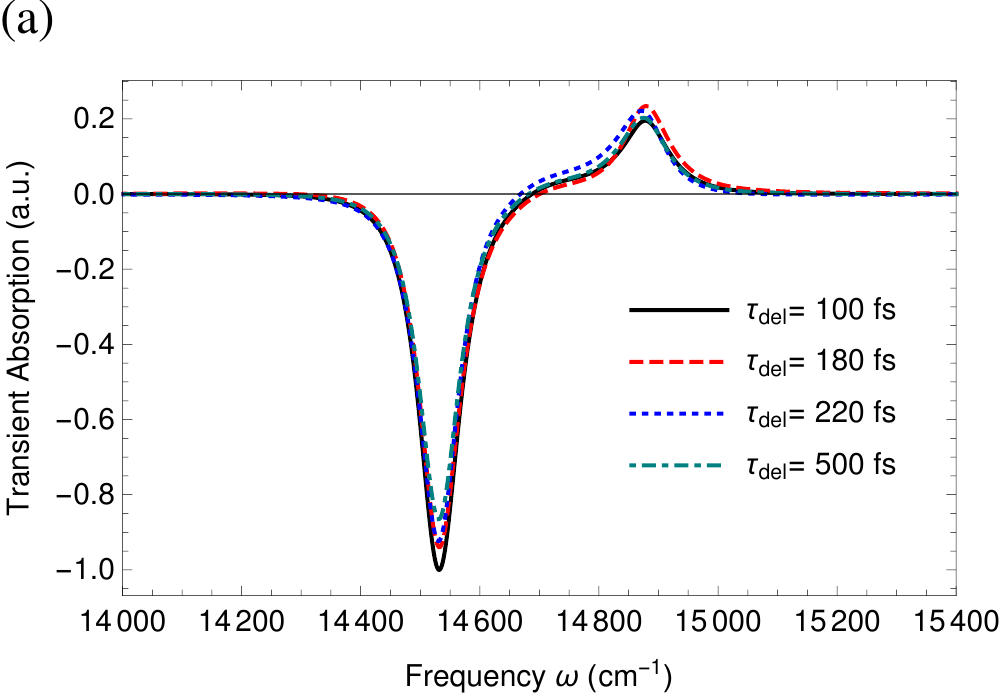}
\includegraphics[width=0.6\textwidth]{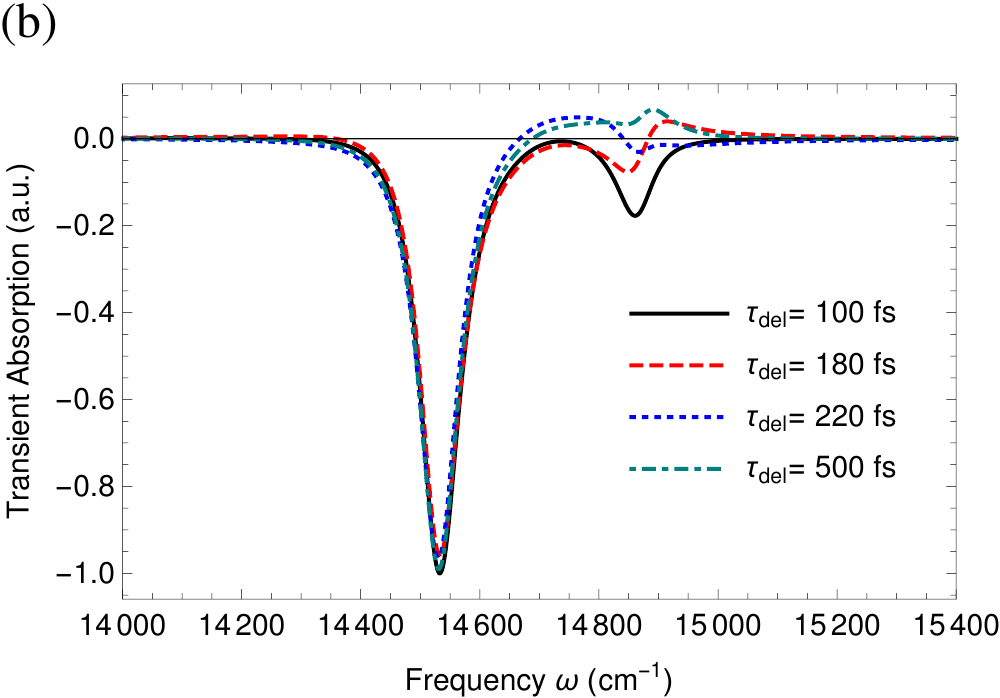}
\caption{
Transient absorption spectra (rotationally and phase averaged) at different delay times $\tau_{\rm del}$. 
(a) $\tau_{\rm pu}=5$~fs; (b) $\tau_{\rm pu}=25$~fs.
}   
\label{figures:transient_spectra_diff_tau_del_small_lambda}
\end{figure}

Fig.~\ref{figures:transient_spectra_diff_tau_del_small_lambda}a shows the transient spectra as a function of delay time $\tau_{\rm del}$ for a broad frequency-band laser pulse.
The temperature is fixed at $T=300$~K for all calculations.
As in the linear absorption spectra discussed before, the specific energy levels and the dipole moments of the model result in a dark region in the low energy part of the spectra due to the charge transfer state.
The negative values, in particular around the second exciton eigenenergy, are caused by the ground state bleaching (GB) and stimulated emission (SE) contributions to the third order optical response function \cite{Mukamel1995,Nuernberger2015}.
The positive part of the transient absorption due to excited state absorption (ESA) is centered around the highest exciton level.
The system dynamics is reflected in the changes of the transient spectra for increasing delay time $\tau_{\rm del}$.
For short time scales and broad energy pump profiles, the laser excitation results in coherent dynamics between the two pigments $P_{D1}$ and $P_{D2}$. 
The coherent dynamics are resulting in shape changes mainly in the ESA region of the transient spectra, around the highest exciton state (Fig.~\ref{figures:transient_spectra_diff_tau_del_small_lambda}b).
At long time scales (for this model $>500$~fs) the system thermalises towards the Boltzmann equilibrium state and there are no noticeable changes in the shape of the transient spectra.

Next, we analyse the effect of the pump pulse width. 
The short pump laser pulse excitation with $\tau_{\rm pu}=5$~fs exhibits a broadband the frequency domain whereas the pulses with $\tau_{\rm pu}=25$~fs and $\tau_{\rm pu}=50$~fs have a narrower bandwidth centered at the highest exciton level. 
By applying a longer laser pulse excitation one selectively excites the exciton states with energies close to the carrier frequency of the pump pulse. 
This leads to the different behavior of the transient spectra for the different pulses as shown by comparing Fig.~\ref{figures:transient_spectra_diff_tau_del_small_lambda}a and Fig.~\ref{figures:transient_spectra_diff_tau_del_small_lambda}b. 
When applying the broadband pulse, all the exciton states are excited with a certain population that depends on the transition dipole moment.
This allows the further transfer of part of the energy to two exciton states even at small delay times. 
Thus, the transient spectra computed at $\tau_{\rm pu}=5$~fs shows a positive region (ESA) for all delay times (Fig.~\ref{figures:transient_spectra_diff_tau_del_small_lambda}a). 
At the same time, Fig.~\ref{figures:transient_spectra_diff_tau_del_small_lambda}b with the narrow bandwidth pump pulse illustrates the dominance of the negative region at short time scales.
This primarily excited highest exciton state cannot contribute immediately to ESA, it decays to the ground state or spreads between the other two excitons and thus results in a negative contribution in the transient spectra.
This is the case for short delay times ($\tau_{\rm del}=100$~fs), while at later times the third exciton spreads and the population of two-exciton states become possible, resulting in the positive contribution in the ESA region of the transient spectra.  
\begin{figure}[t]
\includegraphics[width=0.6\textwidth]{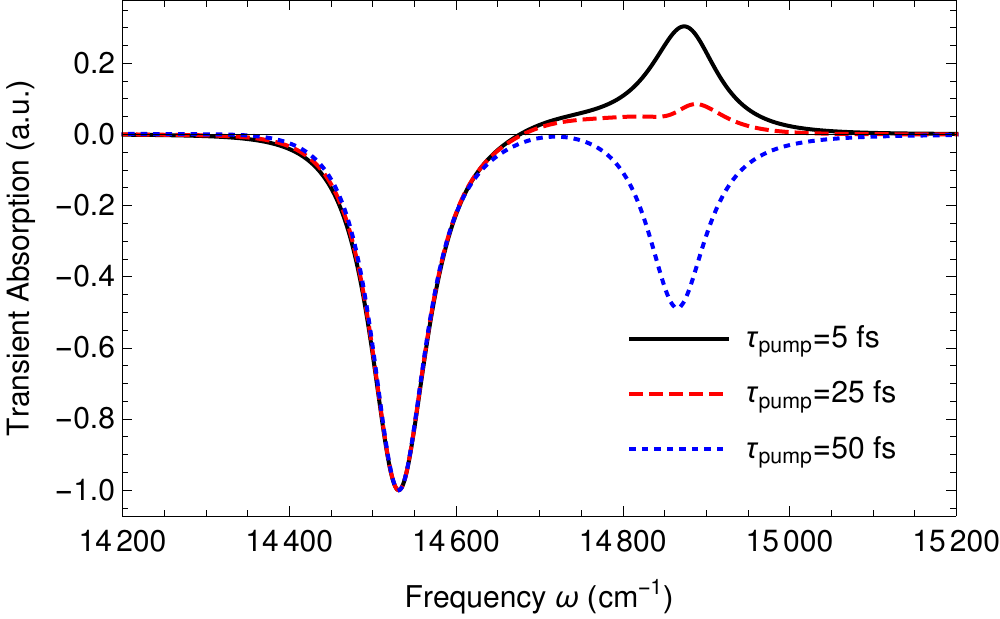}
\caption{
Transient absorption spectra at $\tau_{\rm del}=250$~fs for different pump laser widths and $\lambda=\{20,20,30\}$~cm$^{-1}$.
}   
\label{figures:transient_spectra_diff_pump_width_small_lambda}
\end{figure}

The pump-bandwidth dependence of the transient-absorption spectra is highlighted in Fig.~\ref{figures:transient_spectra_diff_pump_width_small_lambda} for different pulses at the fixed delay time $\tau_{\rm del}=250$~fs.
The narrow bandwidth pulse ($\tau_{\rm pu}=50$~fs) selectively excites the highest exciton state and thus results in vanishing ESA at $250$~fs as it needs some time for redistribution into the lower exciton states. On the contrary broadband frequency pulses with $\tau_{\rm pu}=5$~fs and $\tau_{\rm pu}=25$~fs result in population of the two excitons already at $250$~fs and ESA is not negligible in this case as shown in Fig.~\ref{figures:transient_spectra_diff_pump_width_small_lambda}.

Calculations without any simultaneous excitation of exciton and CT states \cite{Myers2010} (not shown here) result in similar behavior for the TA spectra.

\subsection{Coherence and thermalization time scales}
\label{subsec:pop_dynamics}

\begin{figure}[t]
\includegraphics[width=0.6\textwidth]{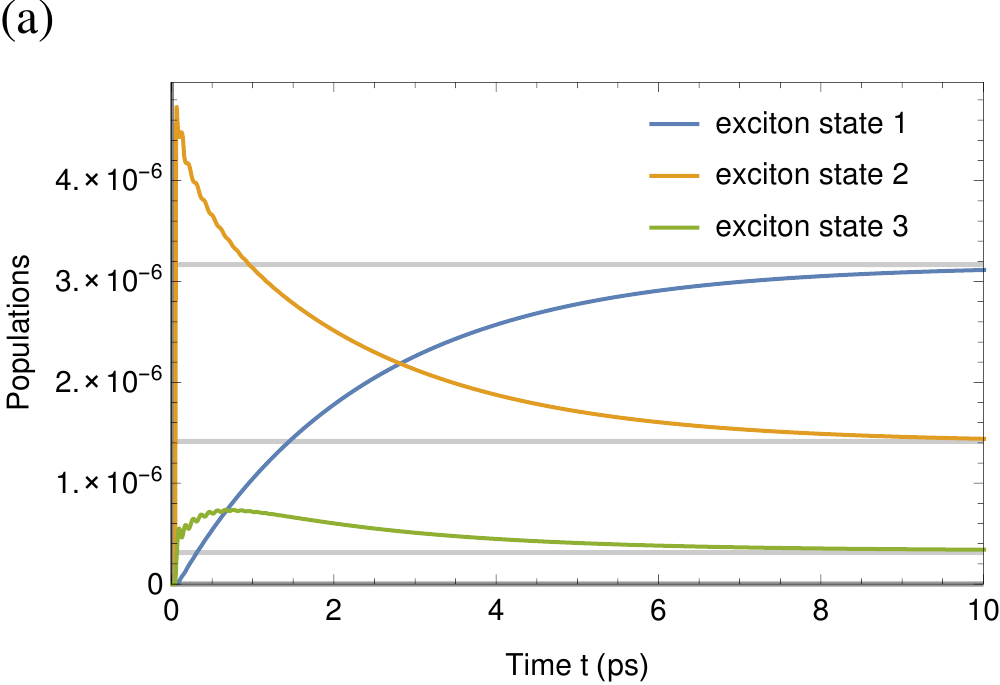}
\includegraphics[width=0.6\textwidth]{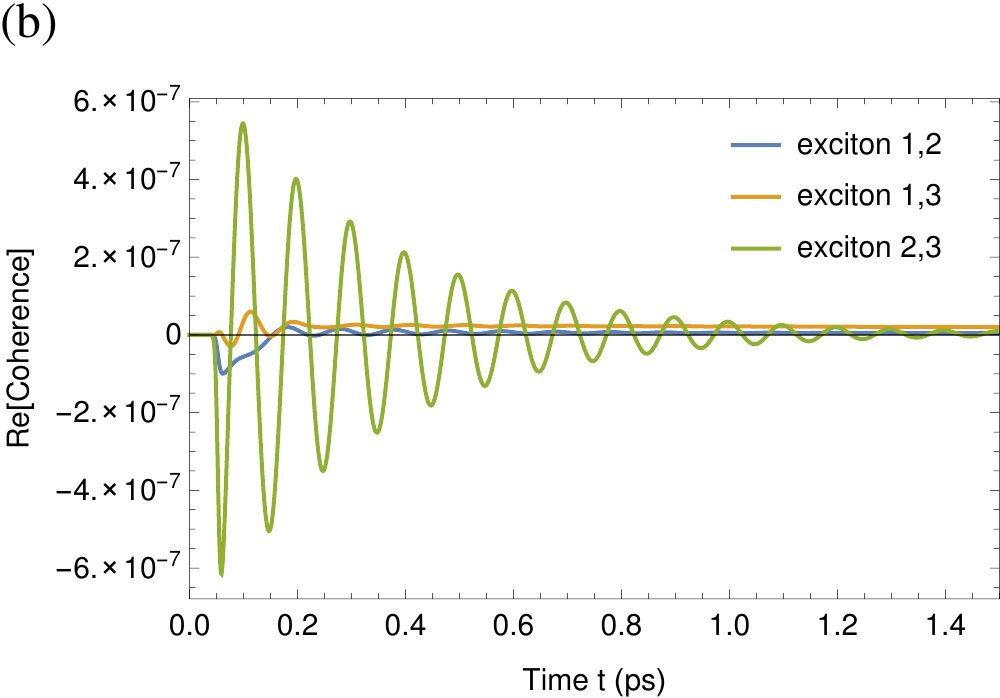}
\caption{
Population dynamics (a) and coherences (b) in exciton representation induced by a $5$~fs pump laser pulse at $T=300$~K for $\lambda=\{20,20,30\}$~cm$^{-1}$.
Exciton states of $H_0$ are numbered from lowest to highest energies.
}   
\label{figures:pop_dynamics_exciton}
\end{figure}
The stimulated emission part of the absorptive signal tracks the time-dependent movement of excitations towards the thermal state, while the ground state bleaching contribution does not encode time-dependent processes. As seen in the previous section, the excited state absorption contribution encodes time-dependent effects due to energy redistribution among the excitons for specific configurations of the pump laser field.

Recently the question of how coherence affects charge transfer dynamics in the reaction center of light harvesting complexes has aroused \cite{Romero2014a, Fuller2014}.
To investigate how closely transient absorption spectroscopy reflects the population dynamics of the system, we calculate the time evolution of system populations for the Hamiltonian $H(t)$  Eq.~(\ref{eq:h-full}) with reorganization energies $\lambda=\{20,20,30\}$~cm$^{-1}$.
Fig.~\ref{figures:pop_dynamics_exciton} shows the population dynamics and coherences obtained by the propagation of the density matrix $\hat{\rho}(t)$, Eq.~(\ref{eq:reduced_density_matrix}) with the broad band pump pulse ($5$~fs) centered at $t_{\rm c,pu}=50$~fs.
The laser excitation of the system leads to the depopulation of the ground state (not shown) with accompanying population of the exciton states, enumerated by increasing eigenenergies.
The population of the second exciton state reaches its maximum value approximately  at $t_{\rm c,pu}=50$~fs which corresponds to largest intensity of the pump laser pulse.
The highest relative population of exciton state 2 results from its highest dipole moment.
With increasing time $t\approx 1$~ps to $10$~ps, the population transfers into the exciton state 1, which represents the low energy charge transfer state.
For $t>10$~ps all populations have reached the steady-state values, closely matching the Boltzmann distribution in respect to the eigenenergies of the $H_0$ Hamiltonian 
(horizontal lines in Fig.~\ref{figures:pop_dynamics_exciton}a).
\begin{figure}[t]
\includegraphics[width=0.6\textwidth]{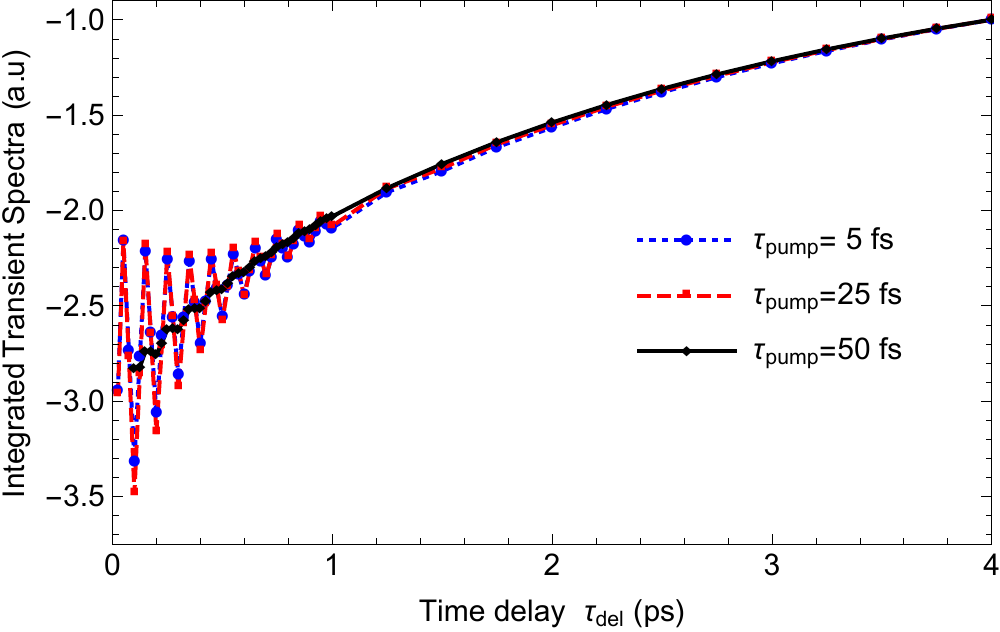}
\caption{ 
Frequency integrated transient absorption spectra versus delay time $\tau_{\rm del}$ for $\lambda=\{20,20,30\}$~cm$^{-1}$.
The overall thermalization process is similar for all considered pump-laser pulse-widths ($\tau_{\rm pu}=\{5,25,50\}$~fs), while the initial coherent beatings are suppressed for the case ($\tau_{\rm pu}=50$~fs).
}   
\label{fig:integrated_transient_spectra_small_lambda}
\end{figure}

The shape and sign changes of the  transient spectra make a simple extraction of time-scales difficult. 
A simpler measure of the overall time-scale in the system is given by the integrated transient absorption spectra, eq.~(\ref{eq:dif_absorption_integrated}), 
displayed in Fig.~\ref{fig:integrated_transient_spectra_small_lambda} for three different pump pulses.
\begin{figure}[t]
\includegraphics[width=0.6\textwidth]{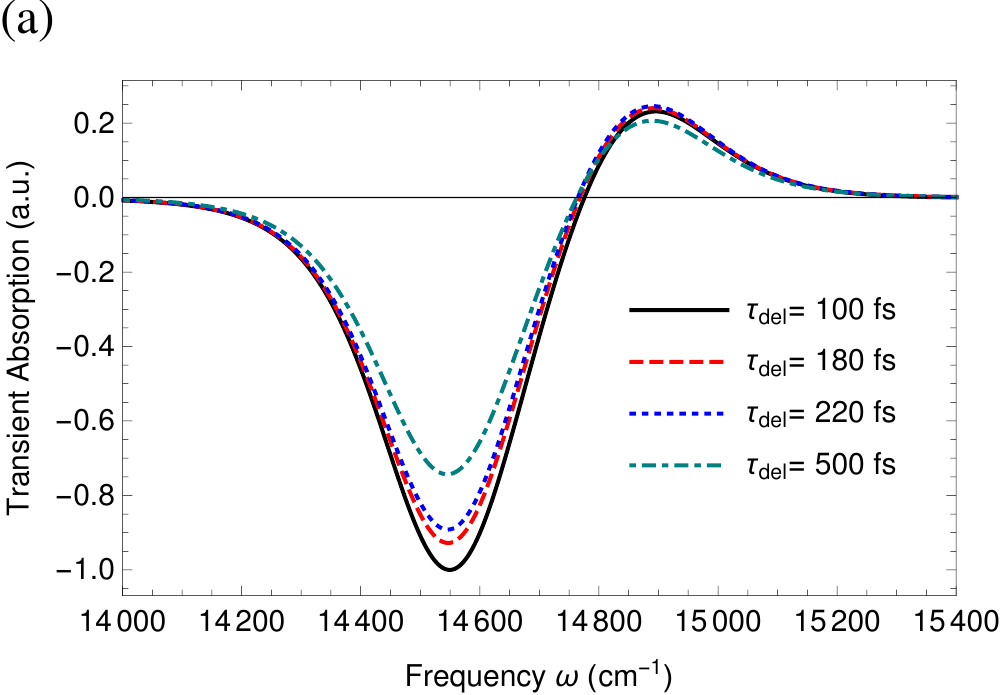}
\includegraphics[width=0.6\textwidth]{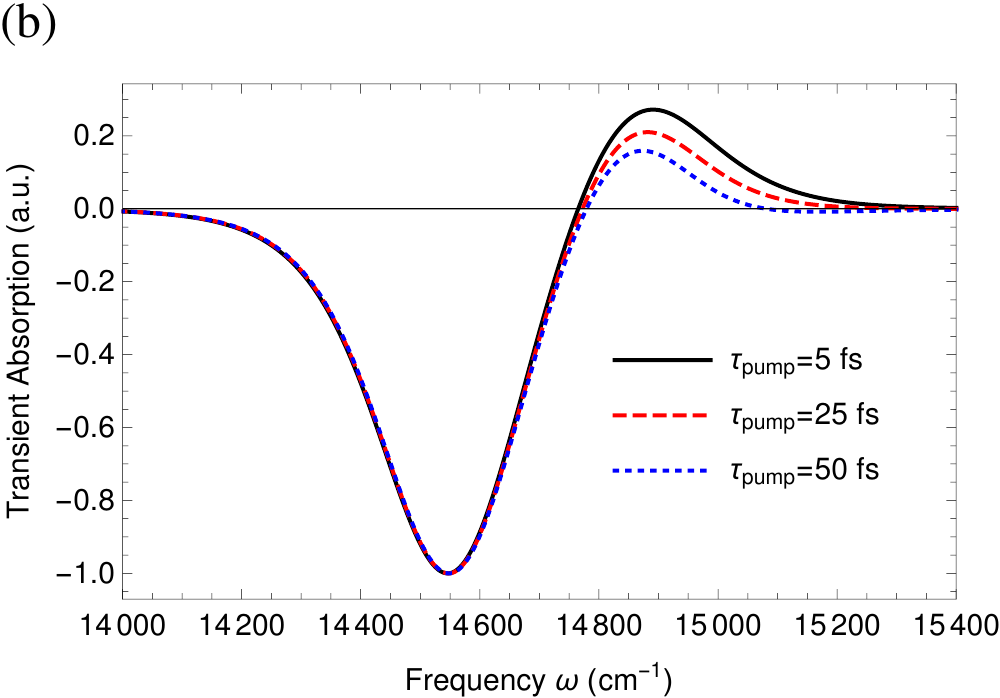}
\caption{
Transient absorption spectra for $\lambda=\{100,100,150\}$~cm$^{-1}$.
(a) narrowband $5$~fs pump pulse, varying delay time.
(b) fixed delay time $\tau_{\rm del}=250$~fs, varying pump pulse width.\label{fig:transient_spectra_large_lambda}
}  
\end{figure}
Despite the integration over the frequency at each delay time, all integrated spectra show oscillatory signals, which are more pronounced for broadband pump pulses (corresponding to shorter pulse widths).
The oscillations reflect the coherent dynamics of the system for times $<1$~ps, and the period correspond to the energy difference between the second and third exciton states, see Fig.~\ref{figures:pop_dynamics_exciton}b.
Besides the oscillations, the population and energy distribution of all exciton states affects the integrated signal.
For longer pulses, the system is excited in a more specific eigenstate and thus coherent beatings are suppressed.
By fitting an exponential function to the frequency integrated TA spectra  as function of delay time, and in addition another exponential function to the envelope of the oscillations, it is possible to infer the two time scales of interest: the decoherence time (0.5~ps) and the thermalization time (2.6~ps).

\begin{figure}[t]
\includegraphics[width=0.6\textwidth]{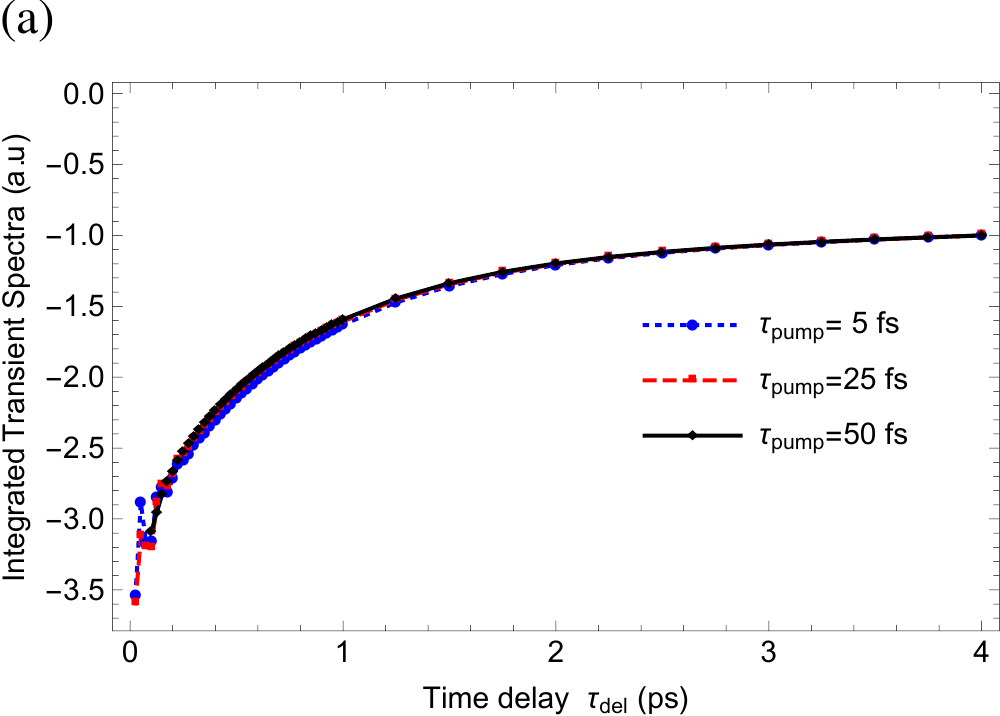}
\includegraphics[width=0.6\textwidth]{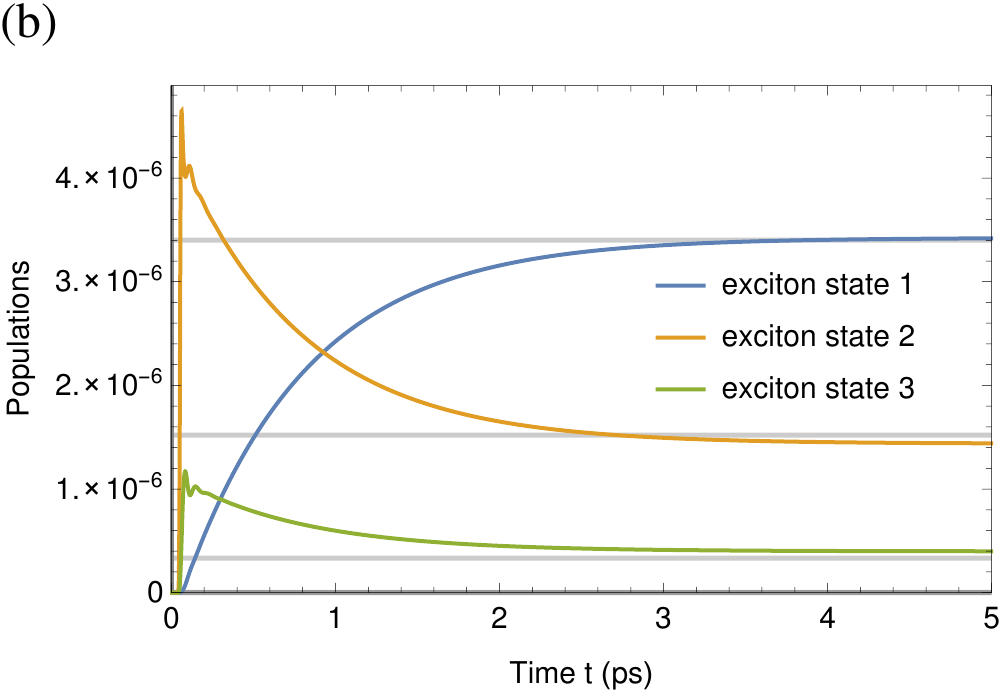}
\caption{ 
(a) Frequency integrated transient absorption spectra versus delay time $\tau_{\rm del}$ for $\lambda=\{100,100,150\}$~cm$^{-1}$. (b) Corresponding population dynamics.
For larger $\lambda$, the thermalization process becomes faster and coherent beatings are absent.
\label{fig:integrated_transient_spectra_large_lambda}
}   
\end{figure}

\subsection{Effect of reorganization energy}

We have presented the results of the charge transfer dynamics within the model system using an unrealistic small reorganization energy for the charge transfer state.
More typical values applied to the photosystem II reaction center are $\lambda\approx 80$~cm$^{-1}$ \cite{Raszewski2005} or $540$~cm$^{-1}$ \cite{Romero2014a,Novoderezhkin2010}.
A smaller coupling results in a slower thermalization and preserves long-lived coherences as compared with those expected from experimental fittings, where the charge transfer between the pigments is assumed to occur in less than $100$~fs, while the thermalisation takes place over the picoseconds time scale \cite{Romero2014a,Novoderezhkin2010,Myers2010}.

An increasing coupling to the environment and the finite time scale of the reorganization process are not contained within the Redfield approach.
Within the numerically exact HEOM method, a larger coupling ($\lambda$) requires to increase the number of auxiliary density operators and to add further Matsubara frequencies for lower temperatures.
Thus, we are restricting the model to reorganization energies below $150$~cm$^{-1}$, which requires a hierarchy depth $K=10$ and $M=2$ Matsubara frequencies.
The main effect of the increasing reorganization energies in the calculated spectra is broadening of the peaks, seen already in the linear absorption spectra Fig.~\ref{fig:abs}. 
Also, time dependent variations of the TA are less visible within the broad peaks in Fig.~\ref{fig:transient_spectra_large_lambda}.
The integrated transient spectra Fig.~\ref{fig:integrated_transient_spectra_large_lambda}a confirms that for larger reorganization energies the electronic coherences between excitonic eigenstates decay much faster than for smaller $\lambda$ ($314$~fs) and also support a fast thermalisation rate of $0.8$~ps  Fig.~\ref{fig:integrated_transient_spectra_large_lambda}b.

\section{Conclusions}
\label{sec:Conclusions}

We have performed a numerically exact HEOM calculation of the transient absorption spectra of a minimal exciton-CT system which models the short time charge transfer dynamics in PSII-RC. 
We have included the effect of the environment and investigated to what extent the model properties are reflected in the spectra.
We discuss how the finite laser pulse strongly affects the excited state absorption, which has been neglected in previous calculations of transient absorption using HEOM\cite{Bai2015}.

From the HEOM calculation, we find that aggregated experimentally accessible quantities such as the integrated transient absorption signal reflects the excitonic properties such as coherences and thermalization time scales.
The time-propagation covering the picoseconds time-scale required to build on the efficient implementation of HEOM on graphics processing units (GPU-HEOM) \cite{Kreisbeck2014}.
Here, we extended the GPU-HEOM method to include a time-dependent Hamiltonian due to the time varying laser fields.

With the full density matrix at hand, we perform the direct comparison of coherence and relaxation time-scales of the optical signal and the corresponding state populations and coherences.
This analysis is not possible in rate-equation based interpretations of transient absorption spectra, where only eigenstate populations after pulse excitations are considered \cite{Romero2014a}.

The time-dependent propagation HEOM method is not restricted to Gaussian pulses, but could be adapted to arbitrary pulse shapes leading to an increased selectivity of the pathways \cite{Nuernberger2015}.
To directly compare with experimental data for transient and two-dimensional spectra of PSII  \cite{Myers2010} requires to further improve the computational scalability of HEOM.

\begin{acknowledgement}
The work was supported by the North-German Supercomputing Alliance (HLRN) and by the German Research Foundation (DFG) grant KR~2889. 
M.R. has received funding from the European Union's Horizon 2020 research and innovation programme under the Marie Sklodowska-Curie grant agreement No 707636. 
\end{acknowledgement}

\end{document}